\documentclass{cupconf}

  \checkfont{eurm10}
  \iffontfound
    \IfFileExists{upmath.sty}
      {\typeout{^^JFound AMS Euler Roman fonts on the system,
                   using the 'upmath' package.^^J}%
       \usepackage{upmath}}
      {\typeout{^^JFound AMS Euler Roman fonts on the system, but you
                   don't seem to have the}%
       \typeout{'upmath' package installed. CUPconf.cls can take advantage
                 of these fonts,^^Jif you use 'upmath' package.^^J}%
      }
  \else
  \fi

  \checkfont{msam10}
  \iffontfound
    \IfFileExists{amssymb.sty}
      {\typeout{^^JFound AMS Symbol fonts on the system, using the
                'amssymb' package.^^J}%
       \usepackage{amssymb}%

      }{}
  \fi

  \IfFileExists{amsbsy.sty}
    {\typeout{^^JFound the 'amsbsy' package on the system, using it.^^J}%

     \usepackage{amsbsy}}
    {}

\newsavebox{\astrutbox}
\sbox{\astrutbox}{\rule[-5pt]{0pt}{20pt}}

\newcommand\araa{{ARA\&A}}%
\newcommand\apj{ApJ}%
\newcommand\apjl{{ApJL}}%
\newcommand\apjs{{ApJS}}%
\newcommand\apss{{Ap\&SS}}%
\newcommand\aap{{A\&A}}%
\newcommand\mnras{{MNRAS}}%
\newcommand\ssr{{Space~Sci.~Rev.}}%
\newcommand\nat{{Nature}}%
\newcommand\grl{{Geophys.~Res.~Lett.}}%
\newcommand\jgr{{J.~Geophys.~Res.}}%
\newcommand\ion[2]{#1$\;${\small family\@Roman{#2}}\relax}%

\def\Rgd{\hbox{$R_{\rm g/d}$}}

\def\Rgd{\hbox{$R_{  gd}$}}

\def\2S12{\hbox{$^{2}{  S }_{1/2}$}}

\def\glong{\hbox{$l^{  II}$}}
\def\glat{\hbox{$b^{  II}$}}

\def\NHI{\hbox{$N {  (H^o )}$}}

\def\nHI{\hbox{$n({  H^o})$}}
\def\nH{\hbox{$n_{  H}$}}

\def\nHII{\hbox{$n({  H^+})$}}

\def\nHeI{\hbox{$n({  He^o})$}}

\def\cmtwo{\hbox{cm$^{-2}$}}

\def\cc{\hbox{cm$^{-3}$}}

\def\deeg{\hbox{$^{  o}$}}

\def\ne{\hbox{$n {  (e^- )}$}}
\def\HI{\hbox{${  H^o}$}}

\def\HII{\hbox{${  H^+}$}}

\def\HeI{\hbox{He$^{  o }$}}

\def\kms{\hbox{km s$^{-1}$}}
\def\cmtwo{\hbox{cm$^{-2}$}}

\def\22V{\hbox{d$V^{  101}_{  s}$}}

\title[ISM and Interplanetary Material]{Galactic Environment of the Sun and Stars:\\
    Interstellar and Interplanetary Material}

\author[Frisch, M{\"u}ller, Zank, and Lopate]{Priscilla C. Frisch and Hans R. M{\"u}ller \and Gary P. Zank and C. Lopate }

\affiliation{$^1$University of Chicago, Chicago, IL\\[\affilskip]
    $^2$Bartol Research Institute, University of Delaware, Newark, DE\\[\affilskip]
    $^3$IGPP, University of California, Riverside, CA\\[\affilskip]
    $^4$University of Chicago, Chicago, IL}

\begin{document}

\maketitle

\begin{abstract}
Interstellar material surrounding an extrasolar planetary system
interacts with the stellar wind to form the stellar astrosphere, and
regulates the properties of the interplanetary medium and cosmic ray
fluxes throughout the system.  Advanced life and civilization
developed on Earth during the time interval when the Sun was immersed
in the vacuum of the Local Bubble and the heliosphere was large, and
probably devoid of most anomalous and galactic cosmic rays.  The Sun
entered an outflow of diffuse cloud material from the Sco-Cen
Association within the past several thousand years.  By analogy with
the Sun and solar system, the Galactic environment of an extrasolar
planetary system must be a key component in understanding the
distribution of systems with stable interplanetary environments, and
inner planets which are shielded by stellar winds from interstellar matter (ISM), such
as might be expected for stable planetary climates.
\end{abstract}

\firstsection 

\section{Introduction}

Our solar system is the best template for understanding the properties
of extrasolar planetary systems.  The interaction between the Sun and
the constituents of its galactic environment regulates the properties
of the interplanetary medium, including the influx of interstellar
matter (ISM) and galactic cosmic rays (GCR) onto planetary
atmospheres.  In the case of the Earth, the evolution of advanced life
occurred during the several million year time period when the Sun was
immersed in the vacuum of the Local Bubble (Frisch and York 1986,
Frisch 1993).
Here we use our understanding of our heliosphere to investigate the
astrospheres around extrasolar planetary systems.\footnote{This paper
is based on the talk presented at the Space Telescope Science
Institute May, 2002 Symposium on the ``Astrophysics of Life''.  See
``Interstellar and Interplanetary Material'', linked to
http://ntweb.stsci.edu/sd/astrophysicsoflife/index.html. }

The heliosphere, or solar wind bubble, is dominated by interstellar
matter (a visualization of the heliosphere is shown in
Fig. \ref{fig:sol1}). Interstellar gas constitutes $\sim$98\% of the
diffuse material in the heliosphere, and the solar wind and
interstellar gas densities are equal near the orbit of Jupiter, beyond
which the ISM density dominates.  The solar wind and photoionization
prevents nearly all ISM from reaching the Earth.  Interstellar ions
and the smallest interstellar dust grains($<$0.1 $\mu$m) are deflected
around the heliosphere. Neutral ISM, however, enters the heliosphere
where it dominates the interplanetary environment throughout most of
the heliosphere, except for the innermost regions where the solar wind
dominates.  Inner and outer planets experience radically different
exposures to raw ISM over the lifetime of a
planetary system. The exposure levels of the Earth to galactic cosmic
rays and raw and processed ISM depends sensitively on heliospheric
properties.

Longstanding theories suggest that interstellar material has
the potential to modify the terrestrial climate. These theories have
recently become less speculative because of the improved understanding
of cosmic ray modulation in a time-varying heliosphere and the
relation between cosmic rays fluxes and atmospheric electricity and
tropospheric cloud cover (Section \ref{sec:climate}). Extrasolar
planetary systems are surrounded by astrospheres formed by the
interaction between stellar winds and interstellar material.  In turn,
these astrospheres modulate the entrance and transport of galactic
cosmic rays, anomalous cosmic rays, neutral interstellar (IS) atoms,
and IS dust into and within the planetary system. Planet habitability
has been evaluated in terms of atmospheric chemistry and energy budget
(see other papers in this volume). However by analogy with the solar
system, an historically stable astrosphere may also be a predictor for
stable planetary climates and thus the conditions which promote
the development of advanced life. It is this relation between the
galactic environment of a star, the stellar astrosphere, and the
properties and prehistory of the interplanetary medium of planetary
systems that are of the greatest interest.

\begin{figure}
\vspace*{4in} 
\caption{{\it The Galactic Environment of the Sun -- Upstream
viewpoint:} Visualization of heliosphere moving through our galactic
neighborhood, based on an MHD simulation of the heliosphere morphology
which includes the relative orientation and ram pressures of the
interstellar and solar wind magnetic fields due to the ecliptic tilt
with respect to the galactic plane (Linde et al. 1998). There is a
north-south asymmetry in the heliosphere from the ecliptic tilt with
respect to the interstellar magnetic field.
The Mach $\sim$1 bow shock around the heliosphere is
apparent, as is the termination shock of the solar wind (the smaller
rounded surface inside of the heliopause where the solar wind
transitions to subsonic). This figure is excerpted from a movie
showing a 3D visualization of the heliosphere and the Milky Way
Galaxy, which can be viewed at http://cs.indiana.edu/$\sim$soljourn.
\label{fig:sol1}}
\end{figure}

\section{Heliosphere and Interstellar Matter}

The heliosphere is the region of space filled by the solar wind, which
is the expanding solar corona.  The solar wind corresponds to a solar
mass loss rate of $\sim$10$^{-14}$ M$_{\rm Sun}$ year$^{-1}$. The
solar wind density decreases with $R^{-2}$ as the solar wind expands,
and the solar wind and interstellar medium pressures are equal at a
plasma contact discontinuity known as the ``heliopause'' (e.g. Axford
1972, Holzer 1989).
The basic properties of the heliosphere are shown in Fig.
\ref{fig:hs} (Zank et al. 1996). At the solar wind termination shock the solar wind
becomes subsonic and the cool supersonic solar wind plasma is
shock-heated to a hot (T$\sim2 \times 10^6$ K) subsonic plasma.
Interstellar neutrals cross the plasma regions with interaction mean
free paths $\sim$100 AU.  If the relative Sun-cloud velocity (26 \kms)
exceeds the fast magnetosonic speed of the surrounding interstellar
cloud, a bow shock will form around the heliosphere.

Solar wind properties vary with the 22-year magnetic activity
cycle of the Sun, 
with the solar magnetic polarity changing every 11 years during the period
of the maximum in solar activity.
During solar minimum, high speed low density solar wind forms in coronal holes at the solar poles ($n(p^+)\sim 2.5$ \cc, velocity $V \sim$ 770 \kms, McComas et al. 2001).
During solar maximum conditions, high speed stream material expands to the
equatorial regions and 
the 1 AU ecliptic solar wind properties are: density $n(p^+)\sim$4--8 \cc, 
velocity $V \sim$350--750 \kms, and magnetic field $B \sim$2
nT (or 20 $\mu$G).  
The activity cycle of the Sun is
known to produce small modifications in the heliosphere over the
11-year solar cycle, with the termination shock moving outwards
$\sim$10 AU in the upwind direction, and outwards by $\sim$40 --
50 AU in the downstream direction during solar minimum.

The Sun is presently in a low density, warm, partially ionized
interstellar cloud with \nH$\sim$0.24 \cc, \ne$\sim$0.1 \cc, and
T$\sim$6,500 K (Slavin and Frisch 2002).  
The upstream direction of the surrounding cloud,
known as the Local Interstellar Cloud (LIC), is towards
\glong=3.3\deeg, \glat=+15.9\deeg\ (in the rest
frame of the Sun) and the relative Sun-LIC velocity is 26.4$\pm$0.5
\kms\ (Witte, private communication).  The LIC upstream direction in 
the local standard of rest (LSR, after removing the solar apex motion) 
is $l$=346\deeg, $b$=--1\deeg\ with
a LIC velocity through the LSR of --15 \kms.
The LIC is a member of a cluster of cloudlets flowing at --17$\pm$5
\kms\ from the LSR upstream direction of \glong=2\deeg,
\glat=--5\deeg\ (Frisch et al. 2002).  The LSR upstream direction is
sensitive to the assumed solar apex motion.\footnote{These quoted
values use a solar apex motion derived from Hipparcos data (Dehnen
Binney 1998).  The basic solar apex motion
yields the LIC LSR upstream direction \glong$\sim$326\deeg, \glat$\sim$+4\deeg\
(Frisch 1995).}

\begin{figure}
\vspace*{4.5in}
\caption{This figure displays the neutral hydrogen density (bottom
panel) and plasma temperature (top panel) of the heliosphere immersed
in the LIC, which has properties T$\sim$6,500 K, \nHI$\sim$0.24 \cc,
\nHII$\sim$0.1 \cc, and an unknown but probably weak magnetic field.
The hydrogen wall is formed by charge exchange coupling between weakly
decelerated and deflected interstellar protons, and interstellar \HI.
\label{fig:hs}}
\end{figure}

The present-day Galactic environment of the Sun yields a highly
asymmetrical heliosphere that is much larger than the planetary
system.  A range of multifluid, Boltzmann-kinetic, and MHD models of
the heliosphere has been developed (see Zank, 1999, for a review).
In the upstream direction, the solar wind termination shock (where the
solar wind becomes subsonic) is at about 75--90 AU.  The heliopause is
located near 140 AU and represents the contact discontinuity between
the solar wind and interstellar plasma component.  The Sun is moving
supersonically with respect to the LIC (sound speed is $\sim$10 \kms),
however a weak interstellar magnetic field ($\sim$3 $\mu$G, fast mode
velocity $\sim$23 \kms) may yield a barely supersonic heliosphere
(M$\sim$1) with a bow shock.  Several heliosphere models place a weak
bow shock at $\sim$250 AU in the upstream direction (see Zank 1999).
For comparison, the planet Pluto is at 39 AU, and the Voyager 1 and
Voyager 2 spacecraft are at 84 AU and 65 AU, respecively.  In the
downstream direction, the termination shock is elongated by a factor of
$\sim$2 compared to the upstream direction.  The north
ecliptic pole points towards the galactic coordinates l=96\deeg,
b=+30\deeg, so the ecliptic plane is inclined by
$\sim$60\deeg\ with respect to the plane of the galaxy.  A pronounced
asymmetry between the northern and southern ecliptic is predicted for
the heliosphere because of this tilt and the LIC upstream direction
(e.g. Linde et al. 1998), combined with the likelihood that the
localized interstellar magnetic field is in the galactic plane (Frisch
1990).

Interstellar plasma piles up against the compressed solar wind in
the outer heliosphere, and charge-coupling between interstellar
\HI\ and interstellar \HII\ produces a low column density
(\NHI$\sim$~3 x 10$^{14}$ \cmtwo),  decelerated ($\delta$V$\sim$8
\kms), heated ($\sim$29,000 K) \HI\ component that is visible as a
redshifted shoulder in the Ly$\alpha$ absorption profile towards
$\alpha$ Cen (Linsky Wood 1996, Gayley et al. 1997, the ``hydrogen
wall'').
Similar pileups of interstellar \HI\ have been detected against the astrospheres
around several nearby cool stars (Section \ref{sec:astrosphere}).

The charged component of the ISM is deflected by the tightly wound
solar wind magnetic field in the heliosheath region.  The smallest
interstellar dust grains($<$0.1 $\mu$m) are also deflected around the
heliopause (Frisch et al. 1999). Neutral ISM, however, enters the heliosphere where it
dominates the interplanetary environment throughout most of the
heliosphere, with the exception of the innermost regions where the
solar wind dominates.

The Voyager 1 and Voyager 2 spacecraft are sending back data from the
frontiers of the outer heliosphere, and future spacecraft may
penetrate interstellar space (e.g., the Interstellar Probe mission,
Liewer and Mewaldt 2000) and provide the first $in~situ$ measurements
of the galactic environment of the Sun.  These spacecraft, and others
(e.g. Ulysses, Galileo, Cassini) have provided a wealth of data which
clearly demonstrate that the ISM dominates the interplanetary
environment throughout most of the solar system and heliosphere.

\section{Historical Variations of the Heliosphere \label{sec:history}}

The Galactic environment of the Sun and stars vary with the motions of
the stars and interstellar clouds through space.  The Sun itself has
been immersed in the vacuum of the Local Bubble (\nHI$<$0.0005 \cc,
\nHII$\sim$0.005 \cc, and T$\sim$10$^6$ K) during the millions of
years over which $homo~sapiens$ developed and civilization emerged
(Frisch and York 1986, Frisch 1993).
The Sun has recently (2,000 -- 10$^5$ years ago) entered an outflow of
diffuse ISM from the Sco-Cen Association (Frisch 1994, Frisch et al. 2002), and is
now surrounded by a warm low density partially ionized cloud.  The Sun
may encounter other possibly denser cloudlets in the flow, with one
possibility being the ``Aql-Oph'' cloudlet that is within 5 pc of the Sun
near the solar apex direction.  
A study of nearby ISM shows 96 interstellar absorption components are seen
towards 60 nearby stars sampling ISM within 30 pc (Frisch et
al. 2002).  Since the nearest stars show $\sim$1
interstellar absorption component per 1.4-1.6 pc, relative Sun-cloud
velocities of 0-32 \kms\ suggest variations in the galactic
environment of the Sun on timescales $<$50,000 years.

The galactic environment of an astrosphere has a striking effect on
the resulting astrosphere.  This is illustrated in Fig. \ref{fig:lbn}
for the heliosphere, which shows the heliosphere properties several
million years ago when the heliosphere was embedded in the Local
Bubble (left), and at some time in the future when it might be
embedded in a cloud with density \nHI=15 \cc\ (but otherwise like the
LIC).  During the time the Sun was embedded in the fully ionized Local
Bubble Plasma, described by T=10$^6$ K, $n$(p$^+$)=0.005 \cc, there
were no interstellar neutrals in the heliosphere, and hence very few
pickup ions or anomalous cosmic rays (very small quantities of each
may have been present from a poorly understood inner source that may
be related to either interplanetary dust or outgassing from planetary
atmospheres).  An increase to n=10 \cc\ for the cloud around the Sun
would contract the heliopause to radius of $\sim$14 AU, increase the
density of neutrals at 1 AU to 2 \cc, and create a Rayleigh-Taylor
unstable heliopause from variable mass loading of solar wind by pickup
ions (Zank \& Frisch 1999).
Models with higher densities (e.g. n=15 \cc, T=3,000 K) show that
planets beyond $\sim$15 AU (Uranus, Neptune, Pluto) will be outside of
the heliosphere for moderate density diffuse clouds, and thus exposed
to raw ISM.  The Sun is predicted to encounter about a dozen giant molecular
clouds, with much higher densities ($>$10$^3$ \cc) over its lifetime
(Talbot \& Newman 1977), but 
encounters with diffuse clouds ($n\sim$10 \cc) will occur more frequently.

\begin{figure}
\vspace*{3.in} 
\caption{Heliosphere predicted for a Sun immersed in the hot Local
Bubble (left) and immersed in a \nH=15 \cc\ diffuse cold cloud
(right). Figure from M{\"u}ller et al. (2002). \label{fig:lbn}}
\end{figure}

\section{Interstellar and Interplanetary Matter \label{sec:ism}}

Components of the interstellar medium which enter the heliosphere from
deep space include neutral gas atoms, larger interstellar dust grains,
and galactic cosmic rays.  The products created by the interactions of
the ISM and solar wind create an ISM-dominated heliosphere.
Fig. \ref{fig:z1} shows an overview of the heliosphere, with the
products of the interaction between the ISM and solar wind identified.

\begin{figure}
\vspace{5.in} 
\caption{Overview of the heliosphere, with
termination shock, heliopause, bow shock, and outer and inner
heliosheath (HS). Some sample plasma (H$^+$), pickup ion (PU Ion),
and solar wind plasma (v$_{\rm HS}$) trajectories are shown, as
well as trajectories of neutral hydrogen (H) coming from the
interstellar medium (H$_{\rm ISM}$) and experiencing charge
exchange (*), and galactic cosmic rays (GCR).  The solar and
interstellar magnetic fields (B) are sketched (based on a plot by J. R. Jokipii).  } 
\label{fig:z1}
\end{figure}

\subsection{High Energy Galactic Cosmic Rays in the Heliosphere}

\begin{figure}
\vspace{3.2in}
\includegraphics{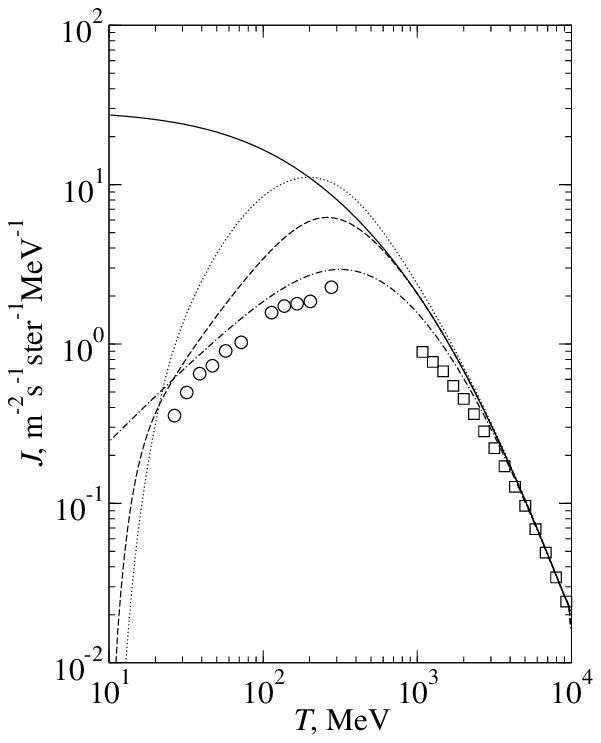}
\vspace{.1in}
\caption{ An example of the modulated cosmic ray spectrum at different
locations in the heliosphere. The dashed line shows the modulated
proton spectrum in the heliosheath ($\theta=0^0$) at 110 AU, the
dash-dotted line is for the supersonic solar wind at 10 AU, and the
dotted line is for the heliotail ($\theta=180^0$) at 650 AU.  The
unmodulated interstellar spectrum is shown as a solid line.
Experimental data from BESS (squares) and IMP8 (circles) are shown for
comparison.  Figure from Florinski et al. (2002).  
(At 10$^2$ MeV, from top to bottom the lines are: solid,
dotted, dashed, dot-dashed.)}
\label{fig:gcr}
\end{figure}

Galactic cosmic rays with energies less than $\sim$100 GeV/nucleon
are modulated by 
the increasingly nonuniform structure of magnetic fields embedded
in the outward flowing solar wind during solar maximum 
(Fig. \ref{fig:gcr}).
The result of this modulation is a well known anti-correlation between
the solar activity cycle and the cosmic ray flux at the Earth's
surface.  The anti-correlation is illustrated in Fig. \ref{fig:1},
which shows neutron monitor counts, from secondary particles produced
by cosmic ray interactions at the top of the atmosphere, versus the
sunspot number.  This anticorrelation reflects variations in the
heliospheric modulation of the galactic cosmic ray flux as a function
of the solar wind magnetic activity.  Anomalous cosmic rays (see
below), formed by accelerated pickup ions, experience modulation in
the heliosphere similar to GCRs.  Most cosmic ray modulation occurs in
the outer part of the heliosphere, so that evidence of CR interactions
on meteorites or planetary surfaces should contain fossil evidence on
the heliosphere radius.  The heliosphere varies with the solar cycle,
as does cosmic ray modulation.  Disorder in the solar wind magnetic
field at sunspot maximum corresponds to an increase in cosmic ray
modulation, although the heliosphere is smaller than at solar minimum.
GCRs are capable of changing the flow pattern of the solar wind and
the surrounding local ISM provided the
particles' coupling to the plasma is sufficiently strong.  
The interstellar cosmic-ray spectra and the diffusion coefficients
and cosmic-ray pressure gradients within the heliosphere are now becoming
better understood (e.g. Ip \& Axford 1985).

\begin{figure}
\vspace*{3.7in}
\caption{Solar cycle modulation of $>$3 GeV galactic cosmic rays: Sunspot
number versus modulated galactic cosmic ray intensity.  
This figure is also available at
http://ulysses.uchicago.edu/NeutronMonitor/neutron\_mon.html along
with related data.}
\label{fig:1}
\end{figure}

\subsection{Raw ISM in the Heliosphere:  \HI, \HeI }

Neutral interstellar H and He atoms enter
and penetrate the solar system, and are ionized by
charge exchange with the solar wind or photoionization.  A weak
interplanetary glow from the fluorescence of solar Ly$\alpha$
radiation off of interstellar \HI, and solar 584 \AA\ radiation off of
interstellar \HeI, led to the discovery of interstellar matter in the
solar system in 1971 (Thomas \& Krassa 1971, Bertaux \& Blamont 1971, Weller \& Meier 1974).  
\HI\ is ionized at $\sim$ 4 AU by charge
exchange with the solar wind and photoionization, while \HeI\
penetrates to $\sim$0.4 AU before becoming photoionized.  The flux of
\HeI\ atoms has been measured directly by Ulysses, yielding values
\nHeI=0.014$\pm$0.002 \cc, temperature 6,500 K, and velocity of 26.4 \kms\ and
and upstream direction \glong=3.3\deeg, \glat=+15.9\deeg\
(Witte et al. 1996 and private communication).  
The first spectral observations of interstellar \HI\ in the solar
system observed a projected velocity --24.1$\pm$2.6 \kms\ during solar
minimum towards the 
direction \glong=16.8\deeg, \glat=+12.3\deeg\ (Adams \& Frisch 1977). 
Correcting this velocity towards the \HeI\ upstream direction
gives a cloud velocity 24.8$\pm$2.6 \kms, in agreement with the \HeI\ velocity
(since during solar minimum radiation pressure and gravity are
approximately equal).  The LSR upstream direction of the LIC is \glong$\sim$346\deeg, \glat$\sim$--1\deeg.

Interstellar \HI\ and \HeI\ behave differently in the heliosphere.
About 20\%--40\% of the \HI\ is lost in the outer heliosheath through
charge-exchange with interstellar \HII, and once in the solar system
the \HI\ trajectory is governed by the relative strengths of the
solar Ly$\alpha$ radiation pressure force and gravity.  Interstellar
\HeI\ passes through the heliosheath unaltered, and the trajectory in
the heliosphere is governed by gravity so that interstellar He is
gravitationally focused downstream of the Sun.  The Earth passes
through the He focusing cone about December 1 of each year.  
The \HeI\ cone density is enhanced at 1 AU by a factor of $\sim$250 over the
value at infinity, but the peak density of the focusing cone is inside
1 AU (Michels et al. 2002).

\subsection{Raw ISM in the Heliosphere: Dust }

\begin{figure}
\vspace*{3.in}
\includegraphics{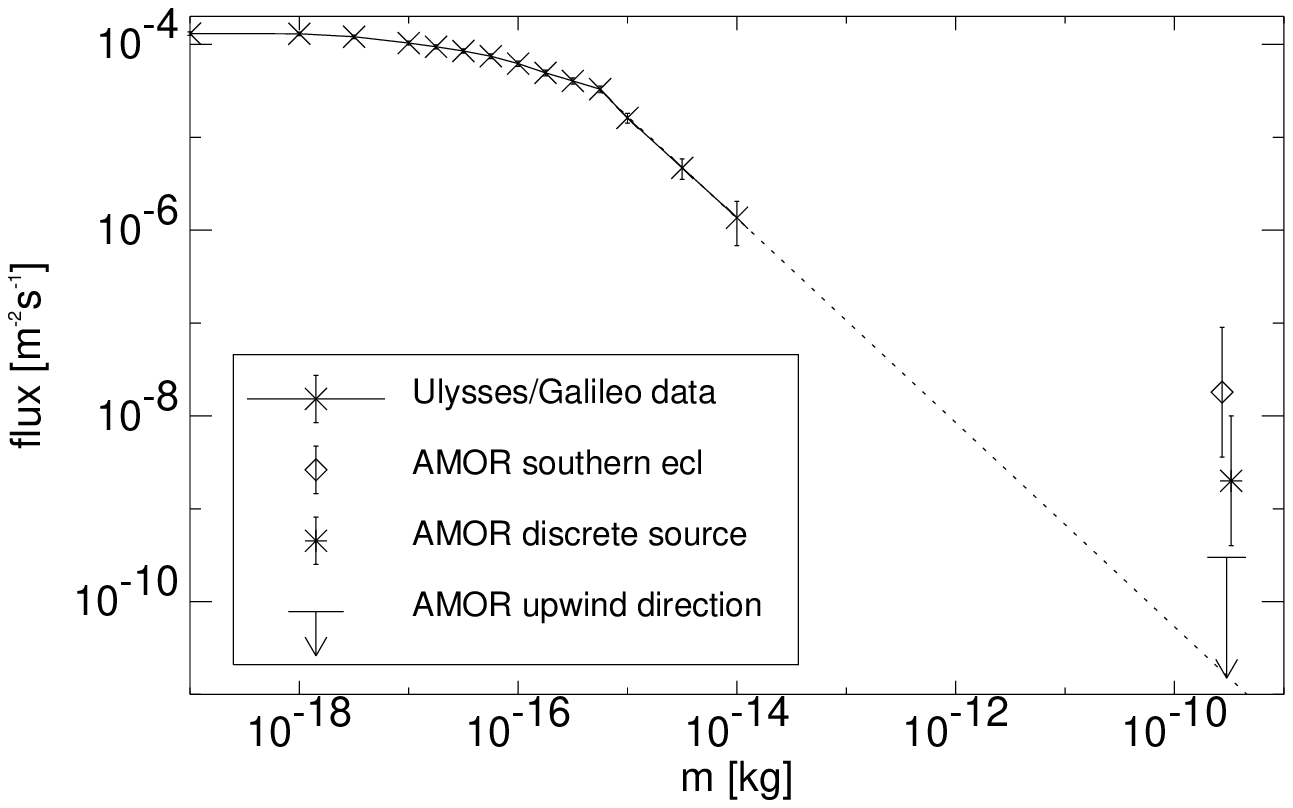}
\caption{Mass flux of interstellar dust grains observed within the
solar system by the Ulysses, Galileo and Cassini spacecraft (Baguhl et
al. 1996, Landgraf et al. 2000).  The AMOR radar data points are of
extrasolar micro-meteorites, and the point source corresponds to a
direction towards $\beta$ Pic (Baggaley 2000).
\label{fig:markus}}
\end{figure}

Interstellar dust grains (ISDG) with radii $>$0.2 $\mu$m enter the
heliosphere and have been detected by instruments on board Ulysses,
Galileo, and Cassini (e.g. Baguhl et al. 1996, Frisch et al. 1999,
Landgraf 2000).
The mass flux distribution of these grains is shown in
Fig. \ref{fig:markus}.  Smaller grains ($< 0.1~\mu$m) are deflected in
the heliosheath region and do not enter the heliosphere.

Large ISDGs (radii $>$0.35 $\mu$m) are focused downstream of the Sun,
in a prominent gravitational focusing cone which is more extensive
than the He focusing cone, extending over 10 AU in the downstream
direction (Landgraf 2000).
Large ISDGs constitute $\sim$30\% of the interplanetary grain
flux with masses $>10^{13}$ gr (or radius$>0.2 ~ \mu$m) at 1 AU (Gruen
\& Landgraf 2000).

ISDGs in the size range comparable to classical dust particles
(0.1-0.2 $\mu$m, charge $\sim$1 eV) show a distribution in the
heliosphere which varies with time because of Lorentz coupling to a solar wind
magnetic field which changes in polarity every 11-year solar cycle.
These positively charged grains alternately are focused and defocused
towards the ecliptic plane.  The 1996 solar minimum corresponded
to a defocusing phase (Landgraf 2000).  

The gas-to-dust mass ratio (\Rgd) in the LIC is \Rgd=125$^{+18}_{-14}$, based on 
comparisons between interstellar dust in the solar system and the
properties for the gas in the LIC, or \Rgd=158 based on missing mass
arguments (Frisch \& Slavin, 2002).

Radar measurements of micrometeorites show sources from outside the
solar system.  Interstellar micrometeorites with masses
$\sim$10$^{-7}$ g are detected by radar observations of the
atmospheric trajectories and velocities (Baggaley 2000, Landgraf et
al. 2000).  A discrete source is seen at the location
of $\beta$ Pic (determined after solar motion is removed).
These observations from the southern hemisphere
also show an enhanced flux from the southern ecliptic.  In the
northern hemisphere, Doppler radar measurements of micrometeorites
provide evidence for a radiant direction towards the Local Bubble
(Meisel et al. 2002).

\subsection{Solar Wind-ISM Interactions Products:  Pickup Ions and Anomalous Cosmic Rays}

Interstellar atoms with first ionization potentials $\gtrsim$13.6 eV enter
and penetrate the solar system, and are ionized by
charge exchange with the solar wind.  The resulting
ions are coupled to the solar wind by the Lorentz force, where they are
observed as a population of pickup ions (PUI, Gloeckler and Geiss
2002).  PUIs of H, He, N, O, and Ne provide a direct sample of
ionization levels in the LIC (Slavin \& Frisch 2002).  PUIs are
accelerated to cosmic ray energies in the region of the termination
shock of the solar wind, forming an anomalous population of cosmic rays 
(Garcia-Munoz et al. 1973, McDonald et al. 1974, Fisk et al. 1974). 
Anomalous cosmic rays, which are ``anomalous'' because of composition and energy,
typically have lower energies than galactic cosmic rays.
The anomalous cosmic ray H, He, N, O, Ne, and Ar populations have an
interstellar origin, and thus provide an additional tracer of the
neutral species in the LIC (Cummings and Stone 2002).
Anomalous cosmic rays with energies $>$1 MeV/nucleon and an
interstellar origin are also found trapped in the radiation belts of
the Earths magnetosphere (e.g. Adams \& Tylka 1993, Mazur et al. 2000).

\section{Astrospheres and Extrasolar Planetary System \label{sec:astrosphere}}

An astrosphere is the stellar wind bubble around a cool star.  Cool
stars with stellar winds will have astrospheres regulated by the
physical properties of the interstellar cloud surrounding each star
(Frisch 1993), and stellar mass loss properties can be inferred from \HI\
Ly$\alpha$ absorption formed in the hydrogen wall region in the
compressed heliosheath gas (Wood et al. 2002).  The nearest star
$\alpha$ Cen AB (1.3 pc)
has a mass loss rate $\sim2$ times greater than the solar value (Wood et al. 2001).
The pileup of interstellar \HI\ in the nose region of astrospheres
surrounding nearby cool stars (e.g. $\alpha$ Cen, $\epsilon$ Eri, 61
CygA, 36 OphAB, 40 Eri A, Gayley et al. 1997, Wood et al. 2002),
indicates that other cool stars have astrospheres which can be modeled
using methodology developed for the heliosphere.

The astrosphere configuration for extrasolar planetary systems will
vary with the individual properties of each system.  The Sun moves
through the local standard of rest with a velocity of V$\sim$13 \kms,
but many cool stars have larger velocities.  Typical diffuse
interstellar clouds move through space with velocities 0--20 \kms\ (or
more), and the dynamical ram pressure ($\sim$V$^2$) may vary by
factors of $\sim$10$^3$, and cause variations in the astrosphere
radius of factors of $>$30.  The result is that inner and outer
planets of extrasolar planetary systems will be exposed to different
amounts of raw interstellar matter over the lifetime of the planetary
system. Frisch (1993) estimated astrosphere radii and historical
galactic environments of $\sim$70 G-stars
within 35 pc of the Sun from the basic Axford-Holzer equation
using the correct stellar dynamics, a solar-like stellar wind, and a
realistic guess for the cloud properties.  However, this primitive
approach can now be improved upon with sophisticated multifluid
astrosphere models (e.g. Zank 1999), improved data from the Hipparcos
catalog, and improved understanding of the nearby ISM.

Astrosphere models, based on self-consistent
algorithms for the coupling of interstellar and secondary neutrals and
ions through charge exchange, predict observable signatures of the
interaction of stellar winds and the ISM.  The interaction products
contain several distinct populations which trace both ISM kinematics
and the underlying donor plasma population.  Comparisons between
predictions of global astrospheric models and Ly$\alpha$ absorption
lines towards nearby cool stars demonstrate that external cool stars
have astrospheres with detectable hydrogen walls.

The modulation of GCRs and ACRs in the heliosphere
indicates that the cosmic ray fluxes in an astrosphere will depend on
the characteristics of the stellar wind interaction with the
surrounding interstellar cloud. Stellar activity cycles give
information on the mass loss from external cool stars. Activity cycles
are observed towards many G-stars, although true solar analogues are
not obvious (e.g. Baliunas \& Soon 1995, Henry et al. 2000).

\begin{figure}
\vspace*{3.5in} 
\includegraphics{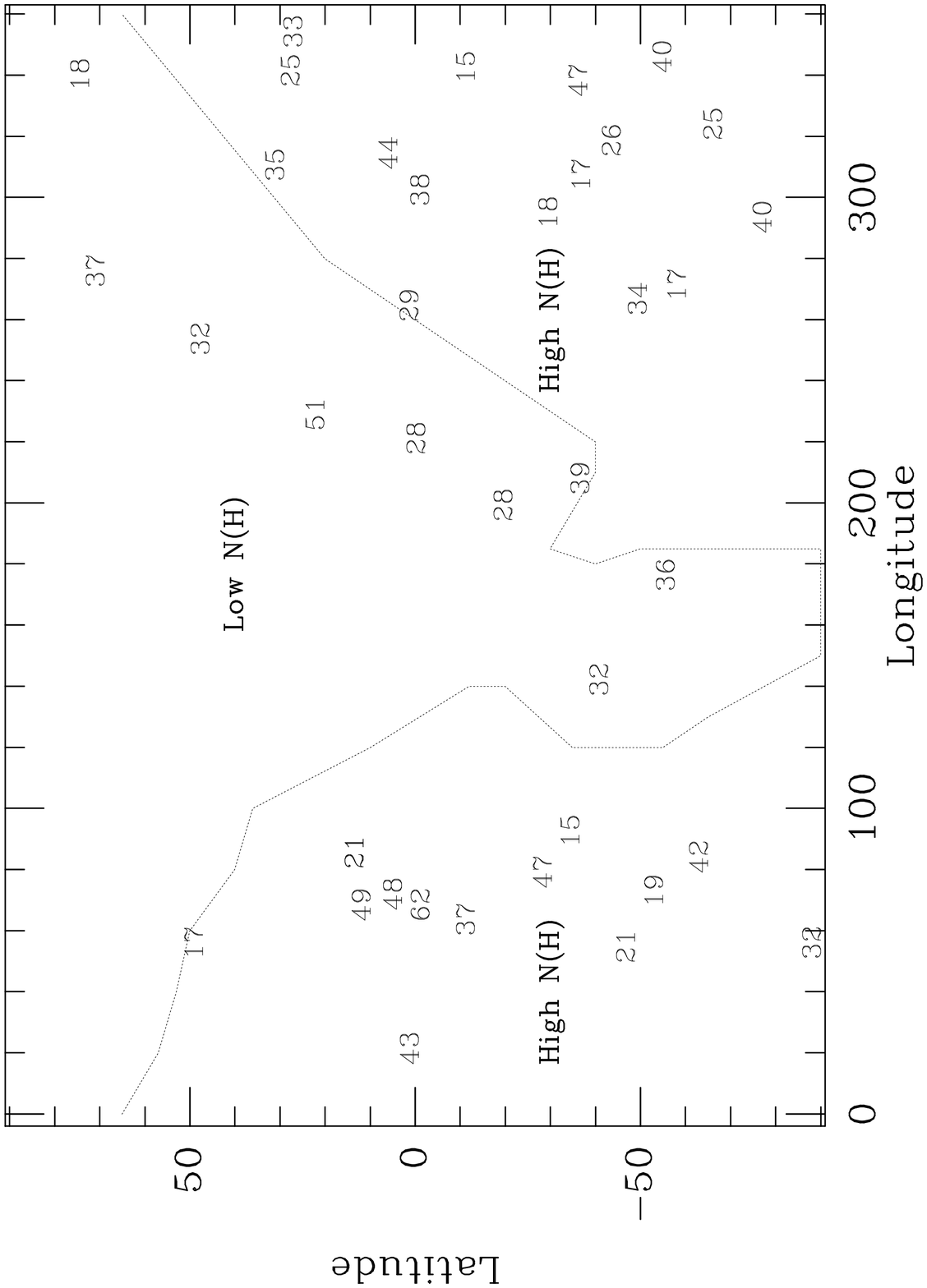} 
\caption{ \small Locations of
$\sim$40 extrasolar planetary systems in galactic longitude and
latitude.  The plotted numbers are the star distance.  The regions
marked ``High N(H)'' show the upstream direction of the cluster of
local interstellar clouds, towards which stars within $\sim$30 pc
are likely to be embedded in a diffuse interstellar cloud. The
direction towards ``Low N(H)'' shows the direction towards the
interior of the Local Bubble or the north pole of Gould's Belt,
towards which stars beyond $\sim$5 pc are more likely to be
embedded in the hot gas of the Local Bubble or high-latitude very
low density ISM.  The N(H) regions are based on Genova et al. (1990).
\label{fig:pla}}
\end{figure}

The galactic positions and distances of $\sim$40 nearby planetary
systems are shown in Fig. \ref{fig:pla}.  The same figure illustrates
the asymmetric distribution of interstellar matter within $\sim$35 pc
of the Sun, with most of the material located in the upstream
direction towards the galactic center (labeled ``High N(H)'') and very
little ISM in the downstream direction (towards the interior of the
Local Bubble, ``Low N(H)'') or near the North Pole (``Low N(H)'').
Stars beyond $\sim$5 pc towards low-N(H) directions are likely to be
embedded in the Local Bubble, while stars within $\sim$40 pc in the
high-N(H) directions are likely to be in diffuse clouds (which may
have densities of up to several particles \cc).  By analogy with the
Sun, the galactic environments of extrasolar planetary systems will
change with time. 

\section{Connections between Astrospheres and Planetary Climates \label{sec:climate}}

Building on the knowledge that the Sun is receding from the
constellation of Orion, an area of active star formation and giant
molecular clouds, Shapley (1921) speculated that the ice ages on Earth
resulted from
a solar encounter with the molecular clouds in Orion.  Since this
earliest speculation, there have been a number of attempts to link
cosmic phenomena and the terrestrial climate.  The investigated
phenomena include (but are not limited to) studies of encounters with
molecular clouds that may be in spiral arms (Thaddeus 1986, Scoville
\& Sanders 1986, Innanen et al. 1978, Begelman \& Rees 1976, McCrea
1975, Talbot \& Newman 1977),
changes in atmosphere chemistry due either to energetic particles from
supernova or the accretion of ISM (Brakenridge 1981, McKay \& Thomas
1978, Butler et al. 1978, Fahr 1968),
nearby supernova (Sonett et al.  1987, Sonett 1997), or
variations in the global electrical circuit or tropospheric cloud
cover from cosmic ray flux variations in the atmosphere (Roble 1991,
Rycroft et al. 2000, Tinsley 2000, Marsh \& Svensmark 2000).


Marsh and Svensmark (2000) presented plausible evidence that a
correlation is present between cosmic ray fluxes and low altitude
($<$3.2 km) cloud cover, which they attribute to cloud condensation
around ionized aerosol particles.  They also argue that low optically
thick clouds cool the climate.  The correlation was observed for low
altitude clouds over the 1980--1995 interval, and the correlation is
dominated by a cosmic ray flux minimum corresponding to the $\sim$1991
solar maximum, using Huancayo neutron counts (cutoff rigidity 13 GeV)
as the cosmic ray monitor.  This correlation, apparently
related to water nucleation on ionized aerosols, provides a possible
mechanism for an astrosphere-climate connection which can be
quantitatively evaluated.

The evolution of advanced life has occurred while the Sun was immersed
in the vacuum of the Local Bubble, and the anomalous cosmic ray
population inside the heliosphere would have nearly vanished and the
enlarged heliosphere would have yielded an effective cosmic ray modulation 
(Mueller et al. 2002). Such a galactic
environment may have promoted stability in the terrestrial climate.

\section{Conclusions}

The evolution of advanced life has occurred during a time when the Sun
was immersed in the vacuum of the Local Bubble, so that the enlarged
heliosphere would have yielded effective modulation of galactic cosmic
rays.  In contrast, an encounter with a modest density diffuse cloud
(n(HI) $\sim$10 \cc) is possible within 10$^4$ -- 10$^5$ years, and
would destabilize the heliosphere and modify cosmic ray fluxes impinging on
the Earth.  The modulation of both galactic and anomalous cosmic rays
by solar wind magnetic fields, and the emerging link between cosmic
ray fluxes and climate forcing, suggests that a stable heliosphere,
and by analogy stable astrospheres, are significant factors in
maintaining climatic stability as is necessary for sustainable
civilization.

The Galactic environment of a star determines interplanetary medium
properties, including the distribution of cosmic rays in the
astrosphere.  How does this affect the ``Astrophysics of Life'', which
is the topic of this conference?  Over the past century many
suggestions have been made regarding Galactic effects on Earth's
climate.  Recent work has demonstrated that the global electrical
circuit is moderated by the cosmic ray flux (Roble 1991), and that,
for instance,
cloud cover in the lower troposphere ($<$3.2 km) correlates with
cosmic ray flux (Marsh \& Svensmark 2002).  The fact which is clear,
however, is that at the present time the solar wind shields the Earth
from most ISM products.  Relatively low fluxes of energetic particles, including galactic
cosmic rays ($>$1 GeV/nucleon) and anomalous cosmic rays ($<$0.5
GeV/nucleon), are able to penetrate to the Earth however.

Simulations which describe the interaction between interstellar clouds
and stellar winds will provide valuable information on the properties
of the astrospheres of extrasolar planetary systems, as well as a
basis for evaluating the interplanetary environment.  Understanding
the historical properties of astrospheres around extrasolar planetary
systems will provide a basis for evaluating the climatic stability on
possible Earth-like extrasolar planets.  The differences in exposure
to raw ISM for inner and outer planets over the planet lifetimes may be significant.

\verb"acknowledgments" PCF would like to thank NASA for research support 
through grants NAG5-8163, NAG5-1105, and NAG5-6405 to the University of Chicago.  
GPZ and HRM acknowledge the partial support of an NSF-DOE grant ATM-0296114 
and NASA grant NAG5-11621.
CL thanks and acknowledges NSF grant 
ATM 99-12341 for providing support for the cosmic ray research.

\end{document}